\documentclass{sig-alternate-05-2015}

\begin{document}

\CopyrightYear{2016} 
\setcopyright{rightsretained} 
\conferenceinfo{SIGIR '16}{July 17-21, 2016, Pisa, Italy} 
\isbn{978-1-4503-4069-4/16/07}
\doi{http://dx.doi.org/10.1145/2911451.2927018}

\clubpenalty=10000 
\widowpenalty = 10000


\title{Learning to Rank Personalized Search Results in Professional Networks}
%
%
%
%
%

\numberofauthors{2} 
%
\author{
%
%
\alignauthor
Viet Ha-Thuc\\
       \affaddr{LinkedIn}\\
       \affaddr{2029 Stierlin Court}\\
       \affaddr{Mountain View, CA, USA}\\
       \email{vhathuc@linkedin.com}
\alignauthor
Shakti Sinha\\
       \affaddr{LinkedIn}\\
       \affaddr{2029 Stierlin Court}\\
       \affaddr{Mountain View, CA, USA}\\
       \email{ssinha@linkedin.com}
}


\maketitle
\begin{abstract}
LinkedIn search is deeply personalized - for the same queries, different searchers expect completely different results. This paper presents our approach to achieving this by mining various data sources available in LinkedIn to infer searchers' intents (such as hiring, job seeking, etc.), as well as extending the concept of homophily to capture the searcher-result similarities on many aspects. Then, learning-to-rank (LTR) is applied to combine these signals with standard search features.
\end{abstract}

%
%

%
%

%
%
\printccsdesc


\section{Introduction}
LinkedIn has evolved over past few years to become a platform containing various professional information sources - including 400+ million member profiles, 6+ million active jobs, millions of professional groups and 18+ million presentations. As the number of sources and their document volumes increase, the problem of serving the right results to fulfill each individual information need becomes increasingly challenging. Moreover, compared to typical information retrieval systems (such as generic Web search), LinkedIn search is deeply personalized. For instance, if a member enters the query ``software engineer'', depending on whether he or she is a recruiter, job seeker or professional content consumer, the member expects to see very different results: software engineers' profiles, software engineer jobs or slideshows on the topic, respectively. Even within a single vertical like Job search, for the query above, if the searcher happens to be an information retrieval expert, he or she is much more likely to be interested in software engineer jobs related to the field rather than software engineer jobs focusing on, say, software QA. For these reasons, we use rich information from the professional network - such as the searchers' identity, profile, network and behavior - to understand their interests and intent, and serve a deeply personalized, relevant search experience. 

With this context, we present our approach to personalize search ranking, leveraging the information available on LinkedIn. At the vertical level, we extend the concept of \textit{homophily} to construct personalized features. Traditionally, the main idea of homophily is that in a social network, people tend to connect or interact with similar people. In the context of LinkedIn, we hypothesize that searchers tend to be interested in results similar or related to them. For example, in People search, searchers are often interested in people in the same or similar industries or companies and people with similar expertise. In Job search, searchers are usually interested in jobs at similar companies, jobs at nearby locations and jobs requiring expertise similar to their own. In particular, for \textit{expertise homophily}, we propose an approach incorporating members' profiles, skill-endorsement graphs and skill co-occurrence patterns to estimate their expertise scores for the skills listed on their profiles, as well as the skills we infer they have. To combine these personalized features with traditional IR features, we apply LTR techniques to learn ranking functions. 

Given results from multiple verticals, a federated model blends them into a single search result page. To personalize the federated model, we mine members' profiles and their recent activities at a large scale to understand their \textit {intents}, such as, hiring intent, job seeking intent or content consuming intent, etc. Then, the federated model is learnt to couple this insight with other signals, including the ones from verticals, to select verticals and to aggregate vertical results into a single ranking that is personally relevant to each of our members.

\section{Search Ranking}
\subsection{Expertise Homophily}
\begin{figure}
\centering
\includegraphics[width=0.28\textwidth]{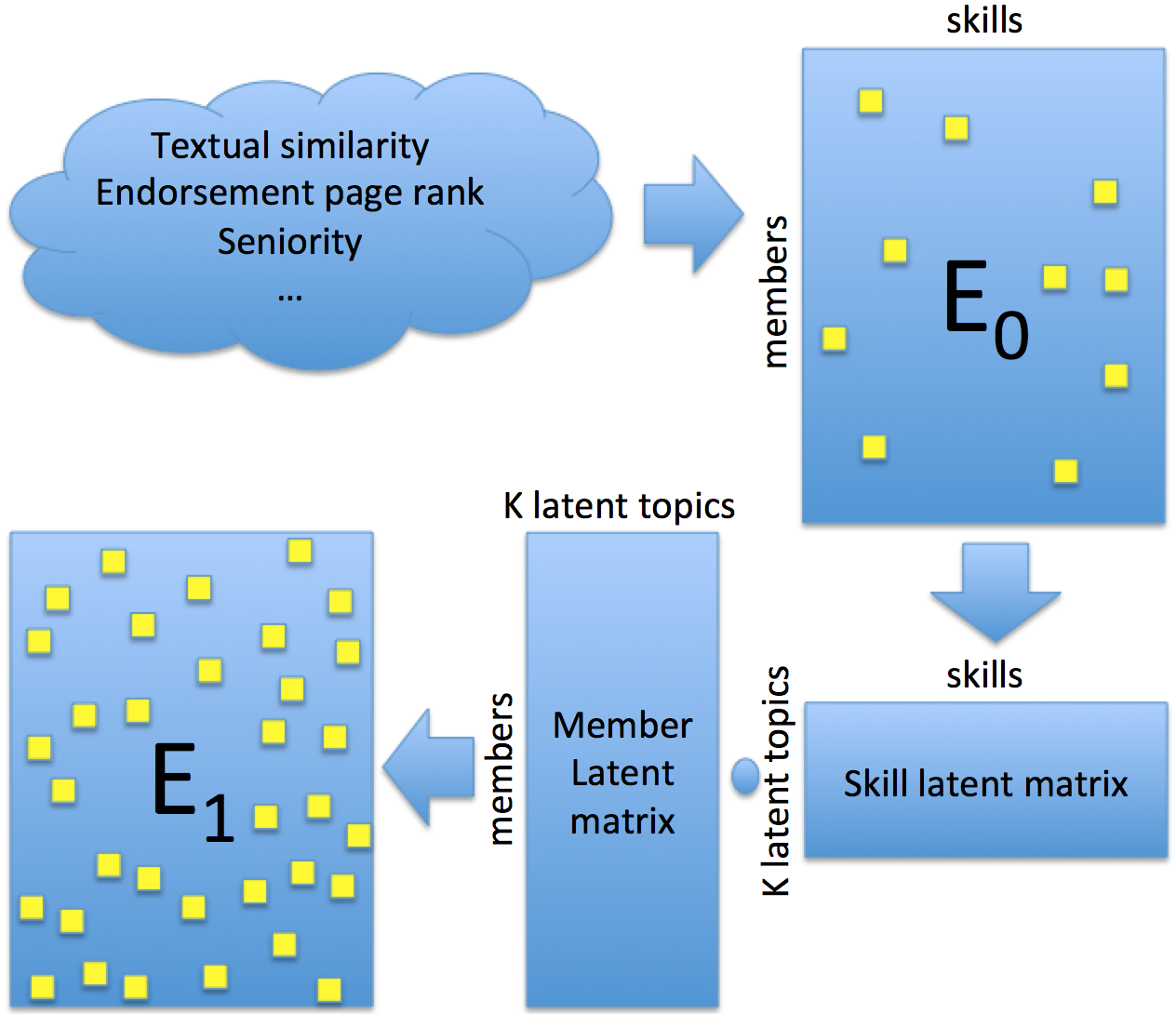}
\caption{Member-skill expertise estimation}
\label{member_expertise}
\end{figure}

On LinkedIn, skills are an integral part of members' profiles that showcase their expertise. A challenge on estimating member expertise is that many members do not explicitly list all skills they have. To overcome this, we employ a two-step approach (Figure \ref{member_expertise}). In the first step, we use a supervised learning algorithm combining various signals on LinkedIn, such as skill-endorsement graph page rank, skill-profile textual similarity, member seniority, etc., to estimate the expertise scores, i.e., $p(expert | member, skill)$. After this step, the expertise matrix ($E_0$) is very sparse since we can be certain about expertise scores only for a small percentage of the pairs. In the second step, we factorize the matrix into member and skill matrices in K-dimensional latent space. Then, we compute the dot-product of the matrices to fill in the ``unknown'' cells. The intuition for this can be illustrated as follows: if a member has ``machine learning'' and ``information retrieval'' skills, based on skill co-occurrence patterns from all of our member base, we could infer that the member is likely to also know ``learning-to-rank''. Interested readers can refer to our recent work \cite{hathuc2016talentsearch,hathuc2015expertisesearch} for more details. The expertise homophily is then estimated by the cosine similarity between the searcher's and results' scores.

\subsection{Learning}
We apply an LTR approach to combine expertise homophily with more traditional search features. In deeply personalized settings like LinkedIn search, editorial judgement does not work well since result relevance strongly depends on individual searchers. Labels derived from search logs containing searchers' actions on results, on the other hand, are personalized by nature. However, the logs have both position and sample selection biases. To avoid position bias, we collect labels from a small traffic fraction that randomizes top-K results. The labeled data is then augmented with \textit{easy negatives}, which are sampled from the tails of rankings to reduce sample bias. The labeled data collection approach is described in \cite{hathuc2015expertisesearch}.  

\section {Federation}
\subsection{Overall Framework}
\begin{figure}
\centering
\includegraphics[width=0.28\textwidth]{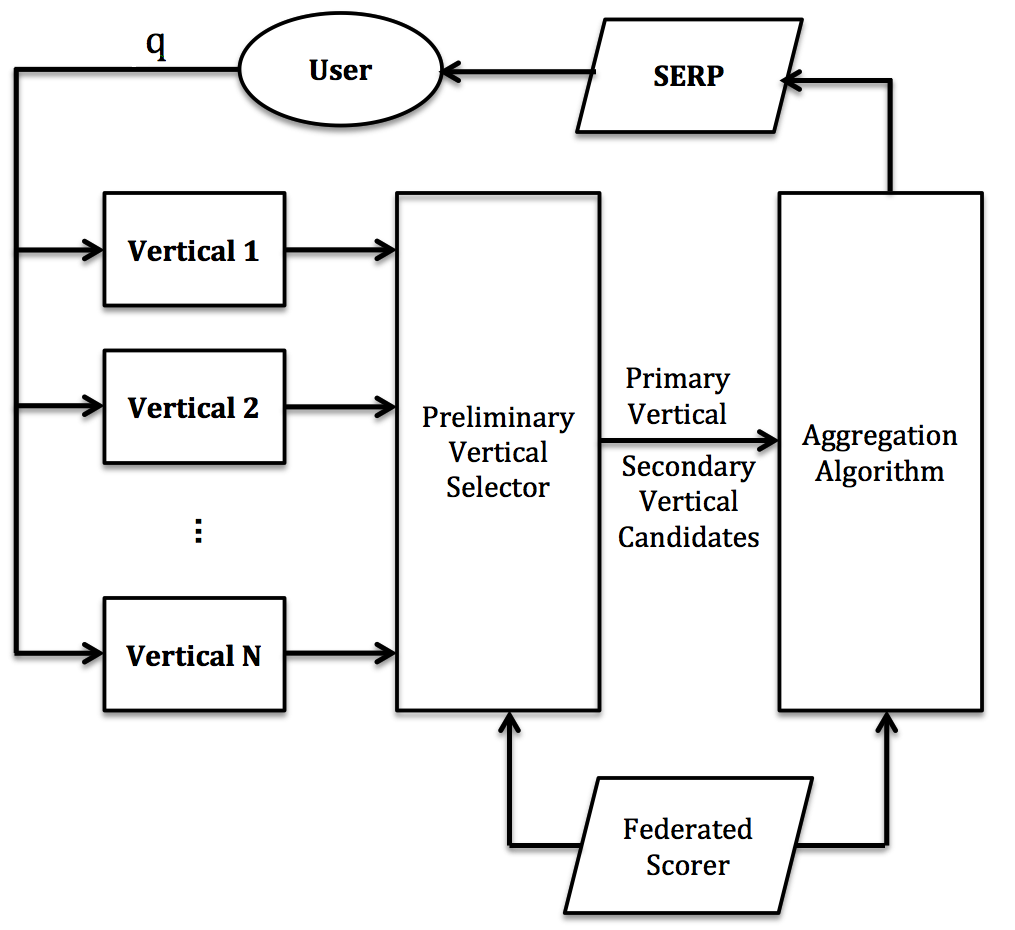}
\caption{Federated Search}
\label{fed_search_framework}
\end{figure}

Given a pair of \textit{(query, searcher)}, the federator selects a primary vertical and a set of secondary verticals, then ranks the primary individual results and the secondary vertical blocks in a single ranked list. The overall framework is described in Figure \ref{fed_search_framework}. When a member issues a query \textit{q}, the query is passed to verticals and triggers the corresponding vertical search engines to get the top \textit{K} results for each. In preliminary vertical selection phase, the federated scorer extracts features and computes a relevance score for each of the verticals. The top vertical is selected as the primary and the rest are selected as \textit{candidates} for secondary verticals. Then, in aggregation phase, these candidates compete with individual results in the primary vertical to form the final ranking. Note that these candidates are not guaranteed to show up in the ranking. Instead, depending on queries, searchers and vertical results, all, some or none of these candidates could be selected. A critical feature used in the federated scorer is searcher intent, which is described in the next subsection. 

\subsection{Searcher Intent}
LinkedIn searchers' information needs constitute a very diverse set of intents. For instance, members actively looking for jobs are likely to be more interested in job results than other verticals. Similarly, members hiring new employees should consider people results to be more important. Given this, we personalize search federation by using the searcher's profile and past activity to infer their \textit{intents}. If a member's job title is recruiter, he is likely to have hiring intent. If a member recently searched for or applied to jobs, he is likely to have job seeking intent. We train machine-learned models combining all of the signals to predict intents for all searchers daily. It is worth noting that a member could have multiple intents at the same time.       

The impact of the intents can vary significantly between different verticals. For example, knowing that a searcher has job seeking intent has much more importance for results that are jobs. We address this variation by constructing \textit{composite features}, capturing both searcher intent and result type. In the example below, the feature is activated only if the searcher has job seeking intent and the result is job. We create combinations of all intents and result categories and learn their weights in our models. In essence, we let the learning algorithm associate evidence with the verticals, and normalize across all verticals from training data. More details can be found in \cite{arya2015federation}.

\[ f = \left\{ 
  \begin{array}{l l}
    1 & \text{if searcher has job seeking intent} \\
      & \text{  AND the result type is job} \\ 
    0 & \text{otherwise}
  \end{array} \right.\]  
  
\section {Conclusions}
We have described how vastly differing intents for the same query arising from different searchers makes LinkedIn search unique and presented the approaches we use to personalize both vertical ranking and federation to deliver a relevant experience to all our searchers.

\bibliographystyle{abbrv}
\bibliography{sigproc}  
%
%

\balancecolumns 
\end{document}